\documentclass[]{spie}  %
\usepackage[]{graphicx}
\usepackage{amsmath,amssymb}

\title{Superconducting quantum dot and the sub-gap states}

\author{Rok \v{Z}itko\supit{a,b}
\skiplinehalf
\supit{a}Jo\v{z}ef Stefan Institute, Jamova 39, SI-1000 Ljubljana,
Slovenia;\\
\supit{b}Faculty  of Mathematics and Physics, University of Ljubljana,
Jadranska 19, SI-1000 Ljubljana, Slovenia
}

\authorinfo{Further author information: (Send correspondence to
R.Z.)\\R.Z..: E-mail: rok.zitko@ijs.si, Telephone: 386 1 477 3571}

\begin{document}
  \maketitle

\begin{abstract}
Quantum dots are nanostructures made of semiconducting materials that
are engineered to hold a small amount of electric charge (a few
electrons) that is controlled by external gate and may hence be
considered as tunable artificial atoms. A quantum dot may be contacted
by conductive leads to become the active part of a single-electron
transistor, a device that is highly conductive only at very specific
gate voltages. In recent years a significant attention has been given
to more complex hybrid devices, in particular
superconductor-semiconductor heterostructures. Here I review the
theoretical and experimental studies of small quantum-dot devices
contacted by one or several superconducting leads. I focus on the
research on the low-lying localized electronic excitations that exist
inside the superconducting gap (Yu-Shiba-Rusinov states) and determine
the transport properties of these devices. The sub-gap states can be
accurately simulated using the numerical renormalization group
technique, often providing full quantitative understanding of the
observed phenomena.
\end{abstract}


\section{INTRODUCTION}
\label{sec:intro}  %

Soon after the discovery of superconductivity it was observed that
magnetic impurity atoms are strongly antagonistic to the
superconducting order: even a small concentration of magnetic dopants
(below one percent) can suppress the critical temperature $T_c$ down
to zero temperature \cite{matthias1958,abrikosov1960,balatsky2006}. At
the same time, superconductivity is remarkably robust to other kinds
of impurities and persists in strongly disordered (even amorphous)
samples \cite{anderson1959sc}. The difference was shown to be related
to the time-reversal invariance, a symmetry with respect to reversal
of the direction of time flow, that is broken in magnets
\cite{anderson1959sc}. Controlled experiments revealed that the
suppression of $T_c$ can be correlated with the spin of the dopant
atoms (rather than their magnetic moment), thus it must be associated
with the exchange interaction between the impurity and the host
\cite{matthias1958}. In addition, the suppression was linear in the
impurity concentration, indicating that this is not a collective
phenomenon, but instead each impurity site acts independently
\cite{matthias1958}. Finally, it was observed that at high impurity
concentrations the moments order magnetically \cite{matthias1958},
implying that the impurities couple with the itinerant electrons of
the superconductor. These three key insights show that the basic
theoretical model for understanding the interplay between
superconductivity and magnetic effects consists of a single magnetic
impurity represented by a point-like quantum mechanical spin which is
exchange coupled with the host electron gas \cite{balatsky2006}. After
the Bardeen-Cooper-Schrieffer theory of superconductivity had been
developped \cite{bardeen1957}, it was shown that a magnetic impurity
modelled as a localized static magnetic moment induces a bound state
inside the superconducting gap
\cite{shiba1968,yu1965,rusinov1969,sakurai1970}. Such excitations are
now known as Yu-Shiba-Rusinov states or Andreev bound states (the
first name will here be used in the following; the two terms are
equivalent when applied to quantum dots which behave as magnetic
impurities). The theory was later extended to incorporate the fact
that magnetic impurities actually have internal dynamics, i.e., that
the impurity spin can flip when host electrons exchange-scatter on the
impurity site \cite{zmha,zmhb,zmhc}. This required a correct
description of the impurity dynamics, a notoriously difficult
theoretical problem known as the Kondo problem
\cite{kondo1964,anderson1967,anderson1969exact1,wilson1975,andrei1981,hewson,glazman1988,
goldhabergordon1998b,cronenwett1998,kouwenhoven2001}.

The Kondo effect (anomalous behavior of normal-state metals at low
temperatures due to magnetic dopants) was experimentally observed in
1930s, and theoretically understood much later in 1960s after noticing
the diverging behavior of high-order perturbation-theory calculations
with decreasing temperatures in a model for a single spin
exchange-coupled to a normal-state metal (the $s-d$ model), despite
the fact that the exchange coupling $J$ is in principle a small
parameter \cite{kondo1964}. Indeed, the problem was later shown to be
non-perturbative in $J$ and required the development of entirely new
theoretical tools to reliably solve it
\cite{anderson1969exact1,wilson1975,krishna1975,andrei1981,rajan1982,andrei1983,affleck1990}.
In some very simplified cases a full analytical solution is possible
for some properties of the model (e.g. for thermodynamics, but not for
the full frequency dependence of spectral functions)
\cite{andrei1981,rajan1982}. A more generally applicable numerical
technique for tackling this problem was develloped in 1970s by K. G.
Wilson \cite{wilson1975,krishna1975,krishna1980a,krishna1980b} and
later refined and generalized
\cite{costi1991,costi1994,bulla1998,oliveira1994,hofstetter2000,peters2006,weichselbaum2007,bulla2008,resolution}.
This approach, now known as the Wilson's numerical renormalization
group (NRG), consists in reformulating the impurity problem as a
one-dimensional continuum problem with a point-like defect. The
continuum is then discretized and the discrete Hamiltonian is
tridiagonalized to obtain a one-dimensional tight-binding chain
representation of the host conduction-band. An appropriate choice of
the discretization scheme leads to exponentially decreasing hopping
constants along the chain \cite{wilson1975}. The impurity is attached
at the beginning of this half-infinite chain. The problem is then
iteratively diagonalized: First an initial block consisting of the
impurity and one or several chain sites is exactly solved by
diagonalisaiton of the relevant Hamiltonian matrix. Then another chain
site is taken into consideration, new Hamiltonian matrices in extended
Hilbert space are constructed, and the resulting eigenproblem is
solved again. Since the Hilbert space grows exponentially with the
number of chain sites included, at some step the states need to be
truncated to the the low-energy subspace. This is actually a good
approximation because the matrix elements relating the low-energy and
high-energy sectors are quickly decaying with increasing energy
difference (the property of energy-scale separation), thus discarding
the high-energy states does not affect the physics on low energy
scales \cite{wilson1975}. The methods allows to calculate the spectrum
of many-body states, thermodynamic properties, expectation values of
all local (neighborhood of the impurity) and some non-local operators,
dynamic quantities, and even the response to quantum quenches (sudden
changes of Hamiltonian parameters) \cite{bulla2008}. Importantly, the
method is applicable to all those situations where the host
Hamiltonian can be reliably approximated by a mean-field decoupling of
the interaction terms. This is notably the case for the reduced
Hamiltonian in the BCS theory of superconductivity.

In early 1990s, the NRG method was for the first time applied to the
problem of magnetic impurities in a BCS superconducting host
\cite{satori1992,sakai1993,yoshioka1998,yoshioka2000}. It confirmed
the presence of Yu-Shiba-Rusinov (YSR) states and for the first time
allowed to very accurately describe their properties. The interest in
this class of problems was renewed in the early 2000s, when it was
realized that a quantum dot embedded between two superconducting
contacts has unusual behavior of the Josephson current (the current
flowing in the absence of voltage difference between two
superconductors with different superconducting phases) in the regime
where it behaves as a magnetic impurity \cite{dam2006}. Furthermore,
experimental techniques were developed to probe single magnetic atoms
on superconductor surfaces using the tip of a scanning tunneling
microscope, giving access to space resolved information
\cite{yazdani1997}. A final impetus to this research field was the
realization that hybrid semiconductor-superconductor devices may host
robust (topologically protected) excitations known as Majorana
zero-modes \cite{mourik2012,deng2012,ruby2015majo}, which are
predicted to be non-Abelian anyons whose braiding corresponds to
unitarily changing the state of the ground-state multiplet of the
system that might lead to an implementation of quantum computers that
are particularly resilient to environment-induced decoherence.

In section II, this contribution presents the basic physics of the YSR
states in quantum dots coupled to superconducting leads. Section III
reviews some experimentally relevant extensions of the model, such as
the effects of additional normal-state tunneling probes and that of
the external magnetic field, high-spin impurities and magnetic
anisotropy effects. Section IV discusses more recent experiments
involving multiple quantum dots and their theoretical interpretation.

\section{YU-SHIBA-RUSINOV STATES}
\label{sec:ysr}

A quantum dot can be electrostatically defined using gate electrodes
deposited on the surface of a semiconductor heterostructure that hosts
a two-dimensional electron (2DEG) gas below its surface
\cite{kouwenhoven1998,kouwenhoven2001rpp}. 2DEG electrons are repelled
by the potential below the electrodes and using suitable patterning it
is possible to obtain a small puddle of electrons coupled to
neighboring 2DEG via tight constrictions: quantum point contacts. A
further gate electrode controls the number of confined electrons.
Alternatively, the electrons can be confined in sections of carbon
nanotubes and semiconductor nanowires, again by suitable shaping the
electrostatic potential using metallic electrodes. The transport
properties of such devices can then be probed by measuring the current
as a function of source-drain bias and gate voltage
\cite{kouwenhovenbook}. One or both of the source/drain electrodes can
be made of a material which becomes superconducting below the critical
temperature $T_C$, giving rise to hybrid superconductor-semiconductor
devices and superconducting quantum dots \cite{hybrid2010}.

\begin{figure}
   \begin{center}
   \begin{tabular}{c}
\includegraphics[height=4cm]{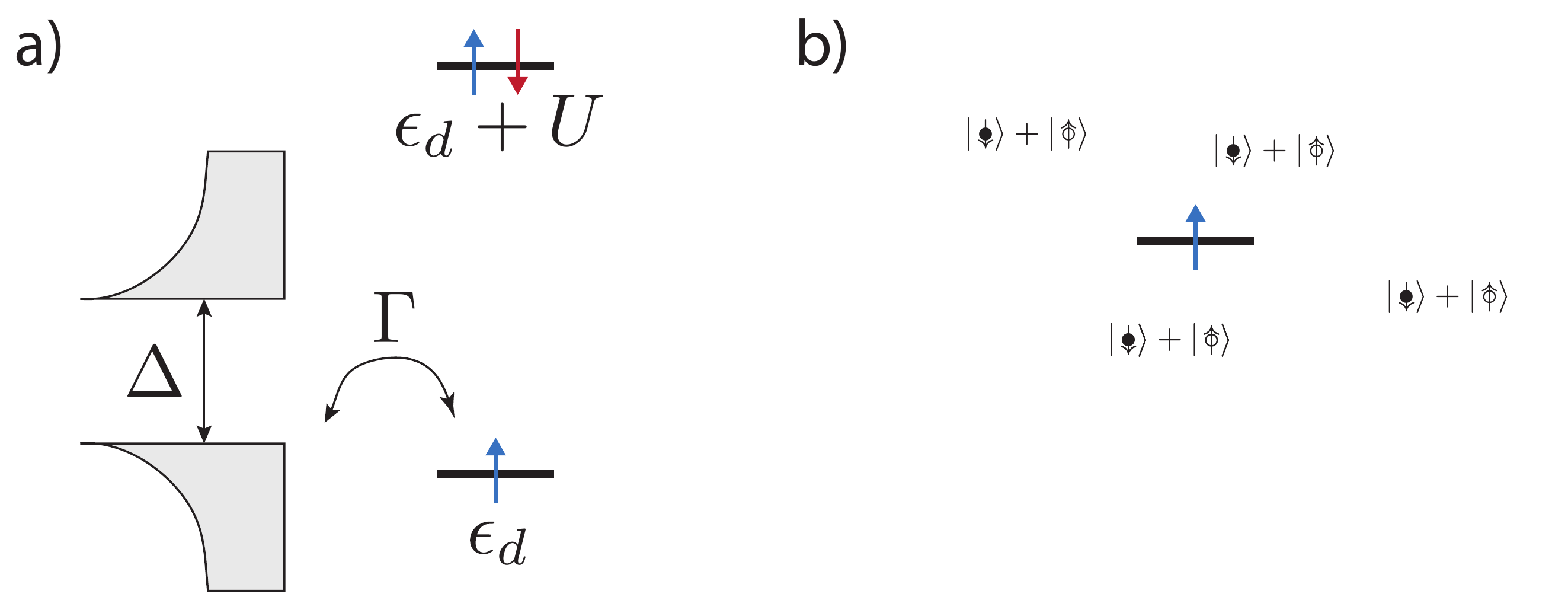}
   \end{tabular}
   \end{center}
   \caption[example]
   { \label{fig:example} (a) Level diagram of the system under
   discussion: an impurity level $\epsilon_d$ with electron-electron
   repulsion $U$ is hybridized with a continuum of excitations in a
   superconductor with a gap $\Delta$ through hybridization with
   strength $\Gamma$. (b) Schematic representation of a
   Yu-Shiba-Rusinov state as the bound state of the impurity spin up
   (blue arrow) and spin-down Bogoliubov quasiparticles
   (superpositions of spin-down electrons and spin-up holes) from the
   superconducting conduction lead.}
   \end{figure}

The simplified description of such a device is obtained by retaining
solely the electron level that is closest to the chemical potential.
(As discussed in the following, this is in fact an excellent
approximation on low energy and temperature scales \cite{pillet2013}.)
The corresponding Hamiltonian then takes the form of the Anderson
impurity model \cite{anderson1967} with superconducting leads
\cite{yeyati1997}:
\begin{equation}
H=H_\mathrm{QD}+\sum_{\alpha} H_{\mathrm{lead},\alpha}
+\sum_\alpha H_\mathrm{c,\alpha}.
\end{equation}
Here
\begin{equation}
H_\mathrm{QD} = \epsilon_d n_d + U n_{d\uparrow} n_{d\downarrow}
\end{equation}
describes the QD energy level $\epsilon_d$ with electron-electron
repulsion $U$, the charge operator is $n_d=n_{d\uparrow} +
n_{d\downarrow}$ with $n_{d\sigma}=d^\dag_\sigma d_\sigma$, where
$d_\sigma$ is the electron annihilation operator. Further,
\begin{equation}
H_{\mathrm{lead},\alpha} = \sum_{k\sigma} \epsilon_k c^\dag_{k\sigma,\alpha}
c_{k\sigma,\alpha} + \sum_{k} \left( \Delta_\alpha c^\dag_{k\uparrow,\alpha}
c^\dag_{-k\downarrow,\alpha} + \mathrm{H.c.} \right),
\end{equation}
is the BCS Hamiltonian for lead $\alpha=1,2,\ldots$, where the
operator $c_{k\sigma,\alpha}$ corresponds to an electron with momentum
$k$, spin $\sigma$ and energy $\epsilon_k$ in lead $\alpha$, while
$\Delta_\alpha=|\Delta_\alpha| \exp(i\phi_\alpha)$ is the BCS order
parameter of lead $\alpha$. Finally,
\begin{equation}
H_\mathrm{c,\alpha} = \sum_{k\sigma}  \left( V_\alpha c^\dag_{k\sigma,\alpha}
d_{\sigma} + \mathrm{H.c.} \right)
\end{equation}
is the coupling between the lead $\alpha$ and the impurity; $V_\alpha$
is the tunnel coupling. The strength of the hybridisation between the dot
and the leads can be conveniently quantified by scalar quantities
$\Gamma_\alpha = \pi \rho_\alpha |V_\alpha|^2$, where $\rho_\alpha$ is
the density of states at the Fermi level in the normal-state of lead
$\alpha$. The parameters can be widely tuned in the experiments,
giving access to very different regimes \cite{jellinggaard2016}. A
schematic representation of the problem is shown in
Fig.~\ref{fig:example}(a).

We first consider the simplest case of a single conduction lead with
$\alpha=1$. The BCS gap $\Delta_1 \equiv \Delta$ can be assumed to be
real with no loss of generality in this case, and we write $\Gamma_1
\equiv \Gamma$.  In the absence of the impurity, this model would have
a spin-singlet ground state (the BCS superconducting wave-function),
separated by $\Delta$ from the onset of the continuum of spin-doublet
excitations (Bogoliubov quasiparticles). For large $U \gtrsim \pi
\Gamma$, the impurity is magnetic \cite{anderson1961} and the Anderson
impurity model in this regime maps \cite{schrieffer1966} onto the
Kondo impurity model with Kondo exchange coupling $\rho J \approx
8\Gamma/\pi U$. It turns out that the ground state of the system then
depends on the ratio of two quantities, $T_K/\Delta$
\cite{zmha,zmhb,balatsky2006,lee2017prb}. $T_K$ is known as the Kondo
temperature and it is given approximately by $T_K \sim U \sqrt{\rho J}
\exp(-1/\rho J)$ \cite{hewson}. This defines the characteristic
temperature and energy scale in the Kondo problem with normal-state
lead: for $T \lesssim T_K$ the effects of the exchange coupling
between the impurity spin and the conduction-band electrons lead to
large effects. In particular, in a normal-state lead the impurity spin
becomes effectively screened for $T \ll T_K$, meaning that a
spatially-extended many-body Kondo state is formed between the
localized spin and the itinerant electrons. This state is a
spin-singlet. In the superconducting case, however, the fermions
required to form the Kondo state are not available to form the Kondo
singlet if $\Delta \gg T_K$. The ground state of the system is then
well approximated by a tensor product of an unperturbed BCS state (a
spin-singlet) and a free impurity spin (a spin-doublet), producing a
many-body ground state that is an over-all spin-doublet. On the
contrary, if $T_K \gg \Delta$, the gap is so small that it hardly
perturbs the Kondo screening and the ground state is essentially the
Kondo singlet state. These considerations imply that with varying
ratio $T_K/\Delta$ an abrupt change of the ground state is expected.
It is indeed directly observed in the NRG calculations: a level
crossing occurs at $T_K/\Delta \approx 0.3$ between a spin-singlet and
a spin-doublet many-particle states. This first-order quantum phase
transition (QPT) between singlet and doublet ground states has a
number of observable consequences, in particular in the spectral
function \cite{balatsky2006,hybrid2010,franke2011}. The singlet and
doublet states are namely connected by the matrix elements that
contribute to the single-particle excitation spectrum, thus the
transition between them is visible in tunneling spectroscopy (see also
below and the subsection~\ref{ssec:probe}). At the QPT, the singlet
and doublet state switch their roles as the ground state and excited
state, respectively, thus spectroscopically this is visible as a
crossing of excitation lines.

\begin{figure}
   \begin{center}
   \begin{tabular}{c}
\includegraphics[height=5.5cm]{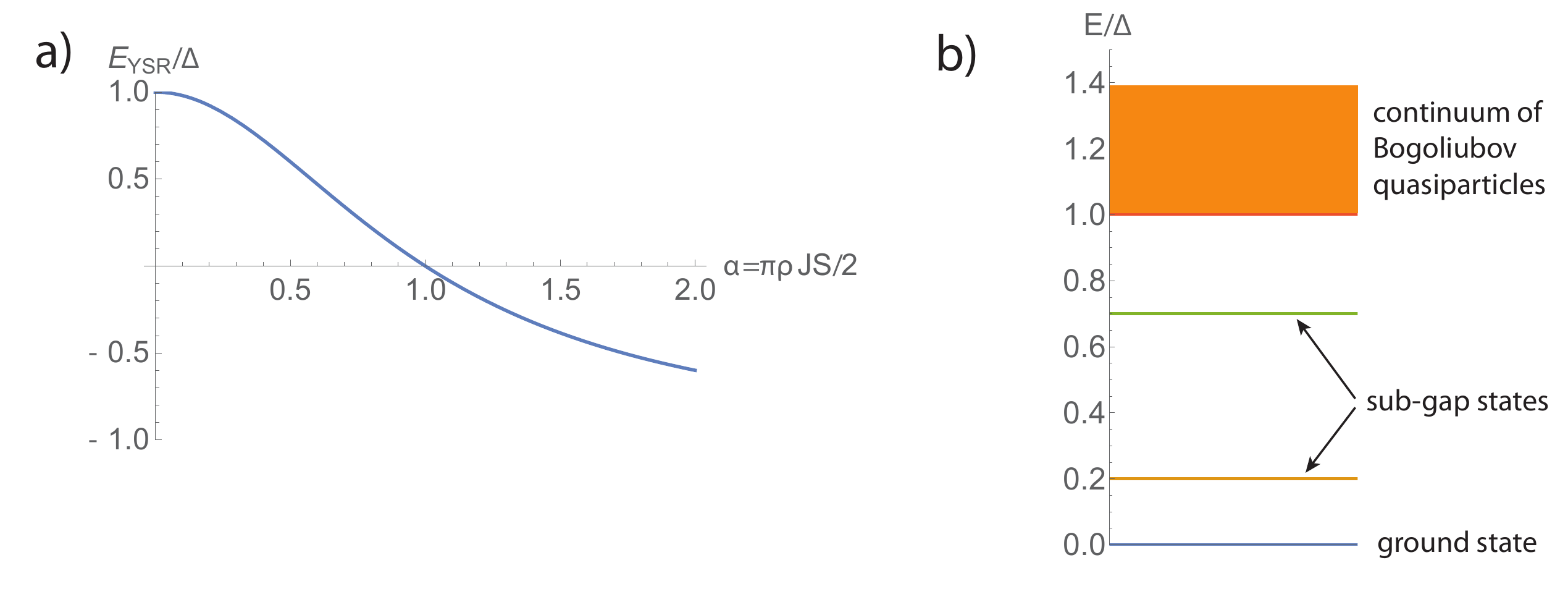}
   \end{tabular}
   \end{center}
   \caption[example]
   { \label{fig2} (a) The binding energy of the Yu-Shiba-Rusinov
   single-particle state in the semi-classical approximation.
   (b) The level diagram for many-particle states in the full quantum
   impurity problem.
   }
   \end{figure}

The original works by Yu, Shiba, and Rusinov addressed a simplified
version of this problem by freezing the impurity spin and neglecting
all spin-flip events \cite{shiba1968,yu1965,rusinov1969}. This is also
known as the ``semi-classical approximation'' that is formally
achieved by taking the $S \to \infty$ and $J \to 0$ limit, so that $JS
= \text{const}$. These authors have shown that a static magnetic
impurity acts as an attractive potential for Bogoliubov quasiparticles
for one spin orientation (that opposite to the polarisation of the
impurity) and repulsive for the other spin species. Since an impurity
is effectively a 1D problem, this implies that a bound state must
exist for attracted Bogoliubov quasiparticles. The binding energy is
given as [see Fig.~\ref{fig2}(a)]:
\begin{equation}
E_\mathrm{YSR} = \Delta \frac{1-\alpha^2}{1+\alpha^2},
\end{equation}
where $\alpha$ is the dimensionless coupling constant, $\alpha=\pi
\rho JS/2$. For low Kondo coupling $J$, the YSR state is located just
below the continuum of Bogoliubov states, but it descends deep down in
the gap region and eventually crosses zero (at $\alpha=1$, i.e.,
$J=2/\pi\rho S$). At this point, a Bogoliubov quasiparticle becomes
trapped at the impurity site\cite{sakurai1970}, thereby ``screening''
the impurity spin, see Fig.~\ref{fig:example}(b). This is actually
highly reminiscent of the behavior in the full quantum problem: apart
from the quantitative details (dependence of the excitation energy on
$J$) and the degeneracy (a doublet of both spin orientations in
quantum case, a single spin polarization in the classical case), the
qualitative behavior is essentially the same. For this reason, the
sub-gap states in the quantum case are also commonly referred to as
YSR states. In general, several such sub-gap can exist at the same
time, see Fig.~\ref{fig2}(b) for illustration.

There are several approaches for characterizing superconducting QD
devices \cite{goffman2000,hybrid2010,rodero2011,bretheau2013}. One
involves measuring the Josephson current, which is a ground-state
property of the system
\cite{vecino2003,choi2004josephson,oguri2004josephson,karrasch2008}.
In particular, it can be shown that the singlet state is associated
with the so-called $0$-junction behavior, where $j \approx j_c \sin
\phi$, with $j_c>0$ being the critical current and
$\phi=\phi_2-\phi_1$ is the phase difference between the SC order
parameters $\Delta_1$ and $\Delta_2$ of the two superconducting leads
contacting the device. The doublet state is, however, associated with
the so-called $\pi$-junction behavior, where $j \approx j_c \sin
(\phi+\pi) \approx -j_c \sin(\phi)$. Such reversals of the Josephson
current at QPTs are indeed experimentally detectable
\cite{dam2006,hybrid2010,maurand2012}. Another commonly used technique
is tunneling spectroscopy, whereby a further (ideally weakly coupled)
electrode at finite voltage is used as a tunneling probe to measure
the density of states (spectral function) at the impurity site
\cite{yeyati1997,pillet2010,buitelaar2002,chang2013}. In addition,
non-equilibrium Andreev transport can also be used
\cite{eichler2007,Deacon:2010jn}.

Experiments have shown that the simple QD devices are often
surprisingly well described by the simplest version of the Anderson
impurity model as introduced in this section. The tunneling spectra
match the calculated spectral functions both on the energy scale
within the gap, as well as somewhat beyond the gap edges in the
continuum \cite{pillet2013}. The agreement between the measured
tunneling spectra and the computed equilibrium spectral functions is
in fact rather surprising, since the dominant transport mechanism at
the lowest temperatures is expected to be that associated with Andreev
scattering that gives particle-hole symmetric spectra (equal spectral
weights at positive and negative bias voltages), while experimentally
one typically observes asymmetric spectra that match the asymmetry
observed in the numerically calculated spectral function
\cite{pillet2013}. This implies the presence of a relaxation
mechanisms (presumably due to phonons) that lead to a fast decay of
excitated states, so that the transport is actually dominated by the
single-electron tunneling. The cross-over between these two transport
regimes has been recently studied through scanning tunneling
spectroscopy of magnetic adsorbates on superconductor surfaces using a
tunneling microscope stabilized to two different temperatures
\cite{ruby2015}. In semiconductor quantum dots, it appears that the
transport is dominated by the single-electron tunneling down to the
lowest temperatures that can be achieved in dilution refrigerators
\cite{ysrdqds}.

Recently a careful quantitative comparison between the theoretical
predictions (as calculated using the NRG) and the experimental
tunneling spectral (measured on InAs nanowires with Al contacts) has
shown excellent agreement over two decades of the $T_K/\Delta$ ratio
\cite{lee2017prb}. This demonstrates the excellent tunability of
modern quantum devices, as well as establishes the NRG results as the
correct solution of the Kondo problem in the case of superconducting
host.

\section{BEYOND THE SIMPLE ANDERSON IMPURITY PROBLEM}
\label{sec:beyond}

Despite the successful description of a quantum dot coupled to
superconducting contacts using the simplest Anderson impurity model
with a single superconducting continuum, there are situation requiring
more involved modeling.

\subsection{Normal-state Tunneling Probes}
\label{ssec:probe}

As previously described, a common technique to characterize quantum
dot devices is tunneling spectroscopy involving an additional weakly
coupled electrode. Ideally, such an electrode is expected to act as a
non-perturbing probe that merely reveals the QD spectral function, but
does not affect the system under study. There are, however, several
caveats. A numerical study has shown that even a relatively weakly
coupled tunneling lead can significantly shift the QPT line in the
phase diagram \cite{zitko2015shiba}. Furthermore, in the case of the
doublet ground state where the impurity remains unscreened, the
continuum electrons in the tunneling probe can act as an additional
screening channel and lead to the emergence of a Kondo resonance in
the tunneling spectrum \cite{zitko2015shiba}. Using a superconducting
lead as a tunneling probe provides no benefit in this regard: in this
case one uses the Bogoliubov states at the edge of the continuum to
probe the QD, again leading to Kondo screening, this time by the
Bogoliubov quasiparticles.

\subsection{External Magnetic Fields}

An external magnetic field has a weak effect on the singlet sub-gap
state, but it directly leads to Zeeman splitting of the doublet
sub-gap state, as observed in experiments \cite{lee2014}. One spin
direction will thus become favored compared to other states, hence the
region in the phase diagram of the model where this state is the
ground state will expand \cite{zitko2015shiba,estrada2018pre}. The
exact behavior strongly depends on the $g$-factors of the impurity and
of the host \cite{willem2017}. The standard theory neglects the
$g$-factor in the host, which is a sensible approximation in system
with very large (absolute value of) impurity $g$-factor, as is for
example the case with InAs nanowires. More generally, the ratio of the
$g$ factors influences the phase diagram of the singlet-doublet
transition as a function of the magnetic field and the $k_B
T_K/\Delta$ ratio \cite{willem2017}.

It is to be noted that in sufficiently strong external magnetic field
a quantum impurity behaves increasingly like a classical, because the
internal spin dynamics become frozen. Similar observation holds for
the effect of the Weiss exchange fields in magnetically ordered
systems. If the N\'eel relaxation time is longer than the typical
experimental times, the mean-field decoupling is a perfectly valid
approximation and theoretical description becomes simple. The most
difficult theoretical situation is thus that of a small number (order
10) of coupled quantum impurities (see section~\ref{sec:multiple}),
each of which is a dynamic object which couples with neighboring
impurities.

Since the sub-gap states are spectroscopically very sharp, it is
possible to very accurately measure the effective $g$-factor of the
quantum dot. This value is reduced from the bare $g$ value by the
renormalization through Kondo exchange coupling, leading to an order
$\rho J$ correction. This provides a further way to extract the Kondo
coupling strength from experimental measurements (although it has not
been put to use so far).

\subsection{High-spin Impurities}

In quantum dots with specific geometry it is possible that several
electron orbitals are partially occupied at the same time, leading to
situations where the electrons form a high-spin many-body state. This
is also the natural state of affairs in most adsorbed magnetic atoms
and molecules on surfaces. In these systems the orbital moment is
partially or totally quenched due to the broken spatial symmetry at
the surface, while the spin degree of freedom persists but experiences
magnetic anisotropy effects due to spin-orbital coupling. See
section~\ref{ssec:aniso}.

These systems can be described using a multi-orbital Anderson or
high-spin Kondo impurity system, similar to the Hamiltonian in section
\ref{sec:ysr}, but with electrons carrying a further orbital quantum
number. As in the single-orbital situation, the impurity spin degrees
of freedom will be screened by the itinerant electrons in a
generalized Kondo effect \cite{cragg1979b,cragg1980}. As a rule, a
single ``channel'' of conduction band electrons (carrying a specific
value of the orbital momentum) can screen a spin-1/2 unit of spin,
thus a spin-$S$ impurity will be fully compensated if it is coupled to
$2S$ different channels. Here, in the context of quantum dots, a
``channel'' means a linear combination from one or several conduction
leads which couples to one of the orbitals of the impurity. To each of
these channels one can assign a corresponding Kondo exchange coupling
constant $J_i$. In turn, each of these defines a different Kondo
temperature $T_{K,i}$. If $J_i$ are unequal, as is usually the case,
the temperatures $T_{K,i}$ will be (exponentially) different. In a
normal-state situation (none of the leads superconducting), the Kondo
screening will then proceed in several stages as the temperature is
reduced \cite{wiel2002}. If some of $T_{K,i}$ are much lower than the
experimental temperature, the moment is partially unscreened and the
impurity remains magnetic. When the leads are superconducting, the
(partially) screened Kondo states will become discrete sub-gap
many-particle states \cite{dqdscaniso}. In the presence of $c \leq 2S$
channels, the original spin-$S$ is screened to $c$ different
spin-$(S-1/2)$ states, $\binom{c}{2}$ spin-$(S-1)$ states,
$\binom{c}{3}$ spin-$(S-3/2)$ states, etc. Some (or most) of these
states will be merged with the continuum and will hence be
non-observable: only those with $T_{K,i} \sim \Delta$ will be
sufficiently inside the gap to play a role. Furthermore, only those
that differ by $\Delta S = \pm 1/2$ from the ground state spin will be
spectroscopically observable in tunneling experiments at zero
temperature, while at finite temperature additional peaks will become
observable, corresponding to the transition between thermally excited
states coupled by the same $\Delta S = \pm 1/2$ selection rule.

\subsection{Magnetic Anisotropy Effects}
\label{ssec:aniso}

In the presence of spin-orbit coupling and reduced spatial symmetry
(notably on surfaces) the spin multiplets will be split
\cite{gatteschi}. This can be described using a Hamiltonian such as
\begin{equation}
H_\mathrm{aniso} = D S_z^2 + E(S_x^2 - S_y^2),
\end{equation}
where $D$ is known as the longitudinal anisotropy and $E$ is the
transverse anisotropy. The directions $xyz$ are here determined as the
principal axes of the magnetic anisotropy tensor. For axial
anisotropy, $D<0$, the absolute value of $S_z$ is maximized and the
impurity behaves as an Ising variable. For planar anisotropy, $D>0$,
the fermion parity determines the resulting state: for integer spin,
the non-magnetic state $S_z=0$ will be the ground state, while for
non-integer spin, the Kramers doublet $S_z=\pm 1/2$ emerges.
For axial anisotropy, the transverse anisotropy is particularly
important: for integer spin, it leads to ``quantum spin tunneling'',
Rabi oscillations between the $S_z=+S$ and $S_z=-S$ states.

When an impurity is coupled to a superconductor, the magnetic
anisotropy splittings are directly visible in tunneling spectroscopy
\cite{moca2008,dqdscaniso}. All spin multiplets are split, and those
connected by $\Delta S_z = \pm 1/2$ (for $E=0$) are spectroscopically
well visible. At finite temperatures, the weights of the spectral
peaks depend on the occupancy of the levels and make it possible to
distinguish the splitting in the ground state from that in the excited
state. This fact allowed to experimentally confirm the presence of
magnetic anisotropic splitting of the $S=1$ state in magnetic
molecules adsorbed on a superconductor surface \cite{Hatter:2016kg}.

\subsection{Thermal effects}

Strictly at zero temperature, and in the absence of environment noise,
the sub-gap states in $s$-wave superconductors should be true $\delta$
peaks with vanishing intrinsic line-width. In reality, there is a
number of mechanisms that lead to a finite line-width. In hybrid
semiconductor-superconductor devices, the induced gap is ``soft'' in
the sense that the density of states inside the gap is not zero, but
there is a instead a finite concentration of quasiparticles.
This mechanism is absent in systems of magnetic adsorbates on thick
superconducting substrates which have ``hard'' gaps. There the
broadening is due solely to thermal effects, most likely linked to
thermally excited vibrational degrees of freedom which are probably
extrinsic to the impurity and rather located in the substrate
\cite{ruby2015}. It is presently unclear whether the electron-electron
interaction may intrinsically lead to finite width of YSR peaks at
finite temperatures. Delicate NRG calculations suggest that this might
be the case \cite{zitko2016thermal}. In magnetically anisotropic Kondo
impurities the presence of a continuum of excitation inside the gap at
finite temperatures has been shown analytically using a perturbative
expansion which is a good approximation in the limit of weak Kondo
coupling constant \cite{lobos2018}.

\subsection{Mesoscopic superconducting islands}

Recently it has become possible to attach a small mesoscopic
superconducting island to a quantum dot \cite{higginbotham2015}. In
this case the superconductor has a finite charging energy,
$H_\mathrm{SC} = E_c (\hat{N}-N)^2$, where $\hat{N}$ is the electron
number operator while $N$ is a parameter which is proportional to the
applied gate voltage. This problem is thus a superconducting
generalization of the Kondo quantum box problem. Since the continuum
becomes interacting in this case, it would seem that the problem is
beyond the scope of the NRG method. One can, however, apply a
technical trick. By introducing a charge operator of the electron box,
one can track the particle transfers between the quantum dot and the
reservoir. The reservoir can then again be considered as
noninteracting and the only price is the introduction of an additional
auxiliary local degree of freedom, which can however be handled in the
NRG calculations \cite{anders2004box}.

\begin{figure}
   \begin{center}
   \begin{tabular}{c}
\includegraphics[height=9cm]{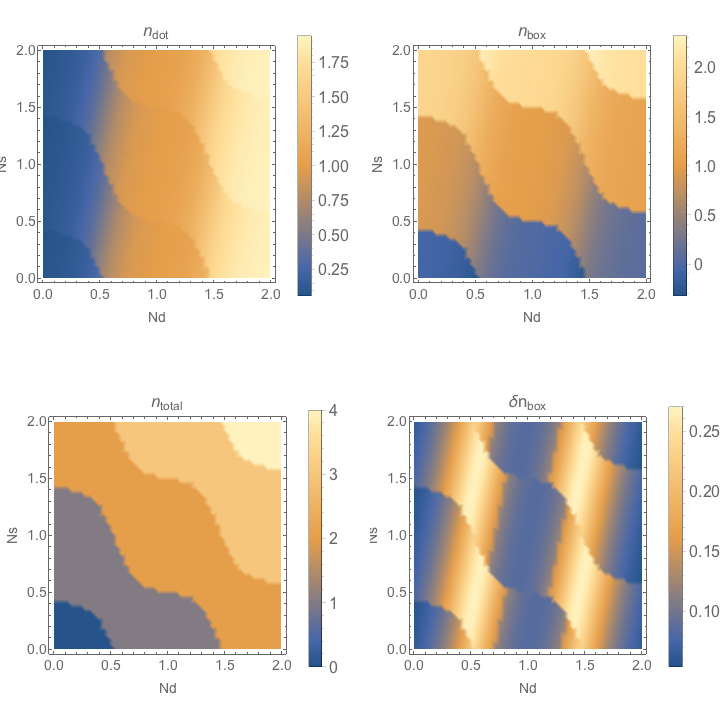}
\includegraphics[height=6cm]{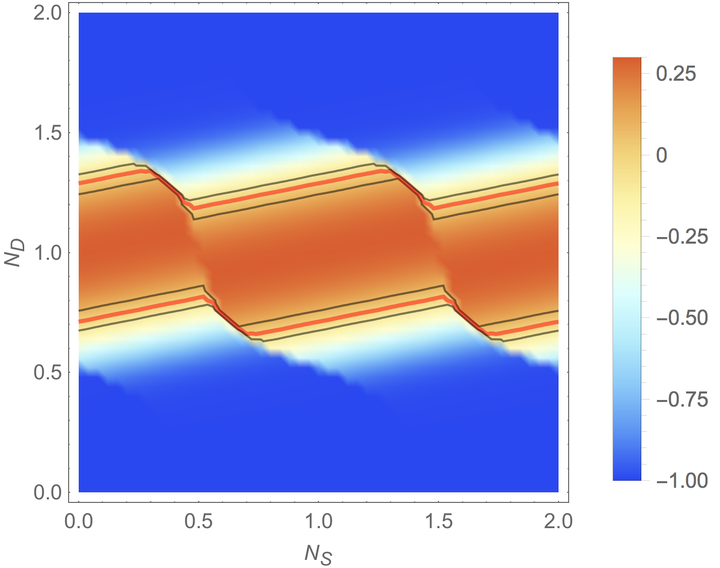}
   \end{tabular}
   \end{center}
   \caption[example]
   { \label{fig:meso} Quantum dot coupled to a small superconducting
   island with sizable charging energy. (a) Electron filling
   (occupancy) of the quantum dot, $n_\mathrm{dot}$, superconducting
   island, $n_\mathrm{box}$, their sum, $n_\mathrm{total}$, and charge
   fluctuations on the superconducting island, $\delta
   n_\mathrm{box}$. The axes correspond to gate voltages applied on
   the dot, $N_D$, and on the reservoir, $N_S$, expressed in
   dimensionless units of charge/occupancy. (b) The energy difference
   between the singlet and doublet states. Positive value corresponds
   to a doublet ground state, negative value to a singlet state. Red
   line is the singlet-doublet transition line and the gray lines
   indicate the width of the transition region. Model parameters are
   $U=1$, $\Gamma=0.1$, $E_c=0.1$, $\Delta=0.01$ in units of the
   superconductor half-bandwidth. }
   \end{figure}

An example is considered in Fig.~\ref{fig:meso}. A quantum dot is
coupled to a superconductor with a moderate value of charging energy
$E_c=U/10$. It is observed that the charging of the superconducting
islands affects the charging of the dot. The charging diagram has the
appearance of the honeycomb and it is not unlike the charging diagram
of double quantum dot structures in its appearance. The modulation of
the charging of the charge box also affects the sub-gap states in the
quantum dot, leading to a periodic pattern. This model may provide a
microscopic interpretation of the results in
Ref.~\cite{higginbotham2015}. In that work an ad-hoc picture in terms
of a fixed-energy bound state was introduced to explain the repetitive
pattern observed in the Coulomb diagrams. Instead, the repetitive
structures find a natural explanation in terms of charging of a
mesoscopic charge reservoir (which here happens to be
superconducting).

\section{MULTIPLE-QUANTUM-DOT DEVICES}
\label{sec:multiple}

Nanostructures constructed out of several quantum dots can be
considered as tunable artificial molecules. They make it possible to
study the inter-impurity interactions induced by tunneling electrons,
such as the formation of molecular-orbital levels via hybridisation of
electron states in individual dots, as well as the exchange coupling
that leads to spin alignment of confined electrons. These effects were
intensely studied in a variety of different realisations of double
quantum dots (DQDs) with normal-state leads
\cite{jeong2001,holleitner2002,craig2004}. More recently, the interest
in the same phenomena resurged in the context of sub-gap
states\cite{yao2014,subgapdqd}, where it is equally interesting to
determine the fate of YSR states in extended systems. The YSR states
are known to hybridize when their wavefunctions overlap, a question of
particular importance for the search of Majorana edge states in
topological superconductors. In addition, there is the super-exchange
interaction that is predicted to lead to different spin states of the
collective many-particle state. Controlled experiments of this type
required the development of tunable devices with multiple (finger)
electrodes. These are now available. They allow independent tuning of
several model parameters, e.g. the on-site potentials on the dots,
coupling to the superconductors as well as that between the dots. At
the same time, the refinement of algorithms and the availability of
powerful computers made it possible to simulate these systems with
increasing accuracy.

In a recent experiment, a DQD was coupled to a single superconductor.
The properties of the device were characterized by sweeping the
potentials of each dot, thereby changing the electron occupancy of
each dot between 0 and 2, and measuring the differential conductance
using an additional weakly-coupled electrode \cite{ysrdqds}. In
particular, the zero-bias differential conductance was mapped as a 2D
diagram in all charge states. For weak coupling to the superconductor,
one obtains results similar to that for normal-state DQDs, the
so-called ``honeycomb diagram'', where the lines of high conductance
correspond to charge degeneracy, i.e., to many-body states with
different occupancy being degenerate in energy. In the superconducting
case, the lines of high conductance correspond to the YSR states
crossing zero frequency in the spectra, but the interpretation is
essentially the same: different YSR sub-gap states become degenerate.
The lines thus delineate the different regimes of singlet or doublet
ground states. As the coupling to the superconductor is increased,
however, the diagrams change their appearance completely: the
singly-occupied charge state (a spin-doublet at weak coupling) on the
dot directly coupled to the superconductor is becoming increasingly
screened. In other words, the YSR singlet drops in energy until it
becomes the ground state in that part of the phase-diagram. This leads
to a characteristic shape and topology of the transition lines. This
behavior was fully captured in theoretical calculations based on a
simple approximation of the superconductor with a single electron
orbital (the so-called ``zero-bandwidth approximation''), as well as in
full NRG simulations \cite{ysrdqds}.

\section{CONCLUSION}

Nanostructures constructed out of quantum dots coupled to
superconductors have in addition to their potential utility as
building blocks of topological quantum computers a more fundamental
scientific value: the superconducting gap namely opens a window to
directly observe and study the competing many-body states that 
manifest as spectroscopically sharp spectral features, while they are
otherwise masked by the continua of excitations in normal-state
systems. As the control of the devices is further increased
(improvements in lithography, harder gaps with reduced residual
density of states in the gap, better control over the gate electrodes,
multi-finger electrode geometries) and the resolution in spectroscopy
is improved (microwave spectroscopy), it is expected that some of the
interesting details that are presently still elusive will eventually
be observed. Some prominent examples are the triplet YSR states in
multiple-quantum-dot structures \cite{subgapdqd,ysrdqds}, as well as
the multiple singlet states in two-channel-Kondo-effect structure
incorporating mesoscopic superconducting islands, indicating the
non-Fermi-liquid nature of the ground state \cite{zitko2017nfl}.

\acknowledgments

R\v{Z} acknowledges the support of the Slovenian Research Agency (ARRS)
under P1-0044 and J1-7259.

\bibliography{article}   %
\bibliographystyle{spiebib}   %

\end{document}